\documentclass[final]{agujournal}

\usepackage{float}
\usepackage{longtable}
\usepackage{bm}
\usepackage{xcolor}

\newcommand{\ssr}{    {Space Sci. Rev.}}
\newcommand{\apj}{    {Astrophys. J.}}
\newcommand{\apjl}{    {Astrophys. J. Lett.}}

\newcommand{\grl}{    {Geophys. Res. Lett.}}
\newcommand{\jgr}{    {J. Geophys. Res.}}

\newcommand{\pre}{    { Phys. Rev. E}}
\newcommand{\prl}{    { Phys. Rev. Lett.}}
\newcommand{\physscr}{{ Physica Scripta}}
\newcommand{\physrep}{{ Physics Reports}}

\newcommand{\blue}{\textcolor{black}}

\journalname{Journal of Geophysical Research}

\begin{document}

%
%

\title{Slow electron holes in the Earth's magnetosheath}

%
%

\authors{Z. I. Shaikh\affil{1}, I. Y. Vasko\affil{1}, I. H. Hutchinson\affil{2}, S. R. Kamaletdinov\affil{3}, J. C. Holmes\affil{4}, D. L. Newman\affil{5,6}, and F. S. Mozer\affil{1}}

\affiliation{1}{Space Sciences Laboratory, University of California at Berkeley, USA}
\affiliation{2}{\blue{Massachusetts Institute of Technology, Cambridge, USA}}
\affiliation{3}{\blue{University of California, Los Angeles, USA}}
\affiliation{4}{Los Alamos National Laboratory, Los Alamos, New Mexico, USA}
\affiliation{5}{Center for Integrated Plasma Studies, University of Colorado, Boulder, CO, USA}
\affiliation{6}{Laboratory for Atmospheric and Space Physics, Boulder, CO, USA}


\correspondingauthor{Zubair I. Shaikh}{zshaikh@berkeley.edu }
\begin{keypoints}
\item \blue{Statistical analysis of 645 solitary waves in the Earth's magnetosheath revealed that 630 of them are electron holes.}

\item The electron holes are associated with quasi-Maxwellian ion velocity distribution functions.

\item The electron hole velocities are comparable with those of the bulk of ions and well below electron thermal speed.

\end{keypoints}

%
%

\begin{abstract}
\justify
We present a statistical analysis of electrostatic solitary waves observed aboard Magnetospheric Multiscale spacecraft in the Earth's magnetosheath. Applying single-spacecraft interferometry to several hundred solitary waves collected \blue{in about two} minute intervals, we show that \blue{almost all of them} have the electrostatic potential of positive polarity and propagate quasi-parallel to the local magnetic field with plasma frame velocities of the order of 100 km/s. The solitary waves have typical \blue{parallel} half-widths from 10 to 100 m that is between 1 and 10 Debye lengths and typical amplitudes of the electrostatic potential from 10 to 200 mV that is between 0.01 and 1\% of local electron temperature. The solitary waves are associated with quasi-Maxwellian ion velocity distribution functions, and their plasma frame velocities are comparable with ion thermal speed and well below electron thermal speed. We argue that the solitary waves \blue{of positive polarity} are slow electron holes and estimate the time scale of their acceleration, which occurs due to interaction with ions, to be of the order of one second. The observation of slow electron holes indicates that their lifetime was shorter than the acceleration time scale. We argue that multi-spacecraft interferometry applied previously to these solitary waves is not applicable because of their too-short spatial scales. The source of the slow electron holes and the role in electron-ion energy exchange remain to be established.

\end{abstract}

\section*{Plain Language Summary}
\justify
Earth's magnetosheath is a highly turbulent medium and an ideal natural laboratory for the analysis of plasma turbulence. Spacecraft measurements showed that high-frequency electric field fluctuations in the Earth's magnetosheath are predominantly electrostatic and consist, particularly, of electrostatic solitary waves with bipolar parallel electric fields. The properties of these electrostatic fluctuations have been largely unaddressed and, moreover, the results of previous studies were inconsistent. In this paper, we present a statistical analysis of electrostatic solitary waves observed aboard Magnetospheric Multiscale in the Earth's magnetosheath. We revealed that most of the solitary waves are Debye-scale structures with the electrostatic potential of positive polarity and typical amplitudes between 0.01 and 1\% of local electron temperature. We demonstrated that the solitary waves must be electron holes,  purely kinetic structures produced in a nonlinear stage of various electron-streaming instabilities. Even more critical is that these structures are {\it slow}; their plasma frame velocities are well below electron thermal speed but coincide with the velocities of the bulk of ions. While the source of electrostatic fluctuations in Earth's magnetosheath could not be revealed, the finding that these fluctuations can be slow implies they can facilitate efficient energy exchange between ions and electrons.

%
%

%



%
%
%

\section{Introduction \label{sec:intro}}

High-resolution electric field measurements aboard modern spacecraft \blue{in numerous regions of the near-Earth space revealed the presence of electrostatic solitary waves, localized electrostatic structures with typically bipolar electric field parallel to local magnetic field} \citep{Mozer15:grl,pickett21:review,Hansel21:jgr}. Prior to the high-resolution measurements, only a signature of solitary waves, broadband electrostatic fluctuations in spectral measurements, was observed \citep[][]{Gurnett76:tail,Gurnett79:pause,Gurnett85}. Broadband electrostatic fluctuations and corresponding solitary waves have been observed in 
the solar wind \citep{Mangeney99,Mozer21:apj}, interplanetary shock waves \citep{Williams05,Wilson07,Wilson10}, lunar environment \citep{Hashimoto10:grl,Malaspina19,Chu&halekas21}, Earth's bow shock \citep{Vasko18:grl,Vasko20:front,Wang21:jgr,Kamaletdinov22:eh}, magnetopause \citep{Cattell:grl02,Graham15}, auroral region \citep{Temerin82,Ergun:grl98,Muschietti:grl99}, inner magnetosphere \citep{Franz:jgr05,Malaspina14,Vasko17:eh_prop,Tong18:grl}, and plasma sheet \citep{Matsumoto94,Cattell05,Lotekar20:jgr}. We have learned that solitary waves can provide efficient electron heating \citep{Vasko18:phpl,Norgren20:jgr,Chu&halekas21} and pitch-angle scattering \citep{Vasko17,Shen21:jgr,Kamaletdinov22:scattering} and also serve as tracers of instabilities not resolvable by particle instruments \citep[][]{Khotyaintsev10:prl,Norgren15:theory,Lotekar20:jgr,Wang20:apjl}. Depending on the space plasma environment, solitary waves were shown to have either predominantly positive or negative polarity \blue{of the electrostatic potential} and were interpreted in terms of electron and ion holes \citep[e.g.,][]{Lotekar20:jgr,Vasko20:front,Wang21:jgr, Wang22:IHs}, purely kinetic structures produced in a nonlinear stage of various electron- and ion-streaming instabilities \citep[e.g.,][]{Roberts67,Omura96,Goldman99:grl,Drake:sci2003,Muschietti08,Che10:grl,Yu21:jgr}. \blue{These structures were named holes, because they are associated with a deficit of electrons or ions trapped within respectively positive or negative solitary wave potential \citep[][]{Roberts67,Schamel86,Mozer18:prl}}. The recent studies have substantially advanced our understanding of solitary wave properties, origin, and effects in various space plasma environments. The only exception is the Earth's magnetosheath, a highly turbulent environment between the Earth's bow shock and magnetopause \citep[e.g.,][]{Narita21:review,Perri21:apj}, where only a few studies of solitary waves have been carried out.


The omnipresence of broadband electric field fluctuations in the Earth's magnetosheath was originally demonstrated by spectral measurements \citep{Rodriguez79:jgr,Mangeney06:angeo}. At frequencies above 100 Hz, these fluctuations are predominantly electrostatic and polarized parallel to the background magnetic field \citep{Rodriguez79:jgr,Mangeney06:angeo}. The high-resolution waveform measurements at a sampling frequency of 4 kHz aboard Geotail spacecraft showed that the broadband fluctuations correspond to electrostatic wave packets and solitary waves with typically bipolar profiles \citep{Kojima97:jgr}. The solitary waves had typical temporal widths of about 1 ms and peak-to-peak electric field amplitudes within about 0.1 mV/m. Using electric field measurements at a sampling frequency of 80 kHz aboard Cluster spacecraft, \cite{pickett03:npg,Pickett05} demonstrated the presence of solitary waves with even smaller temporal widths of 0.025-0.1 ms and peak-to-peak amplitudes of 0.03--0.6 mV/m. Note that neither \cite{Kojima97:jgr} nor \cite{pickett03:npg,Pickett05} could determine the velocity, spatial width, and polarity of the electrostatic potential of the solitary waves. The analysis of polarity and other properties of solitary waves with temporal widths of less than 0.1 ms reported by \cite{pickett03:npg,Pickett05} is still infeasible at present. In turn, \cite{Graham16:jgr} used Cluster electric field measurements at 8 kHz sampling frequency and demonstrated that solitary waves with temporal widths of about 1 ms have plasma frame speeds within a few hundred km/s, spatial widths of a few Debye lengths and {\it positive polarity} of the electrostatic potential. The solitary waves were interpreted in terms of electron holes. Note however that only a few tens of solitary waves were considered by \cite{Graham16:jgr} in the Earth's magnetosheath.

Magnetospheric Multiscale spacecraft have recently provided unique three-dimensional electric field measurements that allowed studying electrostatic solitary waves in the Earth's magnetosphere, including bow shock \citep{Vasko18:grl,Vasko20:front,Wang20:apjl,Wang21:jgr}, magnetopause \citep{Steinvall19:grl}, inner magnetosphere \citep{Tong18:grl,Holmes18:3d} and plasma sheet \citep{Lotekar20:jgr,Norgren20:jgr,Kamaletdinov21:prl}. A study of a few tens of solitary waves in the Earth's magnetosheath was carried out using multi-spacecraft interferometry \citep{Holmes18:jgr}. By correlating seemingly similar waveforms observed aboard different Magnetospheric Multiscale spacecraft, \cite{Holmes18:jgr} obtained solitary wave speeds around 1,000 km/s, spatial widths of a few hundred Debye lengths and {\it negative polarity} of the electrostatic potentials. These estimates are difficult to explain by existing solitary wave theories and also vastly different from solitary wave parameters previously reported in the magnetosheath \citep{Graham16:jgr}. The limited information about solitary waves in the Earth's magnetosheath and inconsistencies between the previous studies have motivated our analysis.

In this paper, we present a detailed analysis of solitary waves observed aboard Magnetospheric Multiscale in the Earth's magnetosheath in the interval considered previously by \cite{Holmes18:jgr}. Instead of multi-spacecraft interferometry, we use single-spacecraft interferometry to estimate the speed and other properties of \blue{645} solitary waves and reveal \blue{that all of them except 15} are actually Debye-scale structures of {\it positive polarity} with plasma frame speeds of the order of 100 km/s. We argue that \blue{multi-spacecraft interferometry is not applicable for the solitary waves} because of their too-short spatial scales. We obtain a statistical distribution of the solitary wave parameters and demonstrate that these structures must be {\it slow electron holes}, whose origin, stability, and other peculiar properties are discussed.

\section{Data and case studies \label{sec2}}


We consider Magnetospheric Multiscale (MMS) observations in the Earth's magnetosheath on November 2, 2016, from 09:31:15 to 09:32:55 UT, when MMS spacecraft were on the dayside close to the magnetopause. We use MMS measurements in burst mode: DC-coupled magnetic field at 128 S/s (samples per second) resolution provided by Digital and Analog Fluxgate Magnetometers \citep{Russell16}, AC-coupled electric field fluctuations and voltage signals of three pairs of voltage-sensitive probes at 8,192 S/s resolution provided by Axial Double Probe \citep{Ergun16} and Spin-Plane Double Probe \citep{Lindqvist16}, magnetic field fluctuations at 8,192 S/s resolution provided by the Search Coil Magnetometer \citep{LeContel16}, electron and ion moments at 30 and 150 ms cadence provided by Fast Plasma Investigation instrument \citep{Pollock16}. The Hot Plasma Composition Analyzer measurements are used to determine the ion content \citep{Young16:ssr}. Voltage signals of the voltage-sensitive probes measured with respect to the spacecraft body are used to compute the electric field. \blue{Two} voltage-sensitive probes are mounted on 14.6 m axial antennas along the spacecraft spin axis, while another four probes are mounted on 60 m antennas in the spacecraft spin plane. The voltage signals are also used for interferometry analysis of the solitary waves, and the corresponding methodology is exhaustively described in the previous study \citep{Wang21:jgr}. Note that the spacecraft spin plane is essentially equivalent to the ecliptic plane or the $xy$ plane of the Geocentric Solar Ecliptic (GSE) coordinate system, while the spin axis is parallel to the $z$-axis of that system.

\begin{figure}
\centering
\includegraphics[width=1\linewidth]{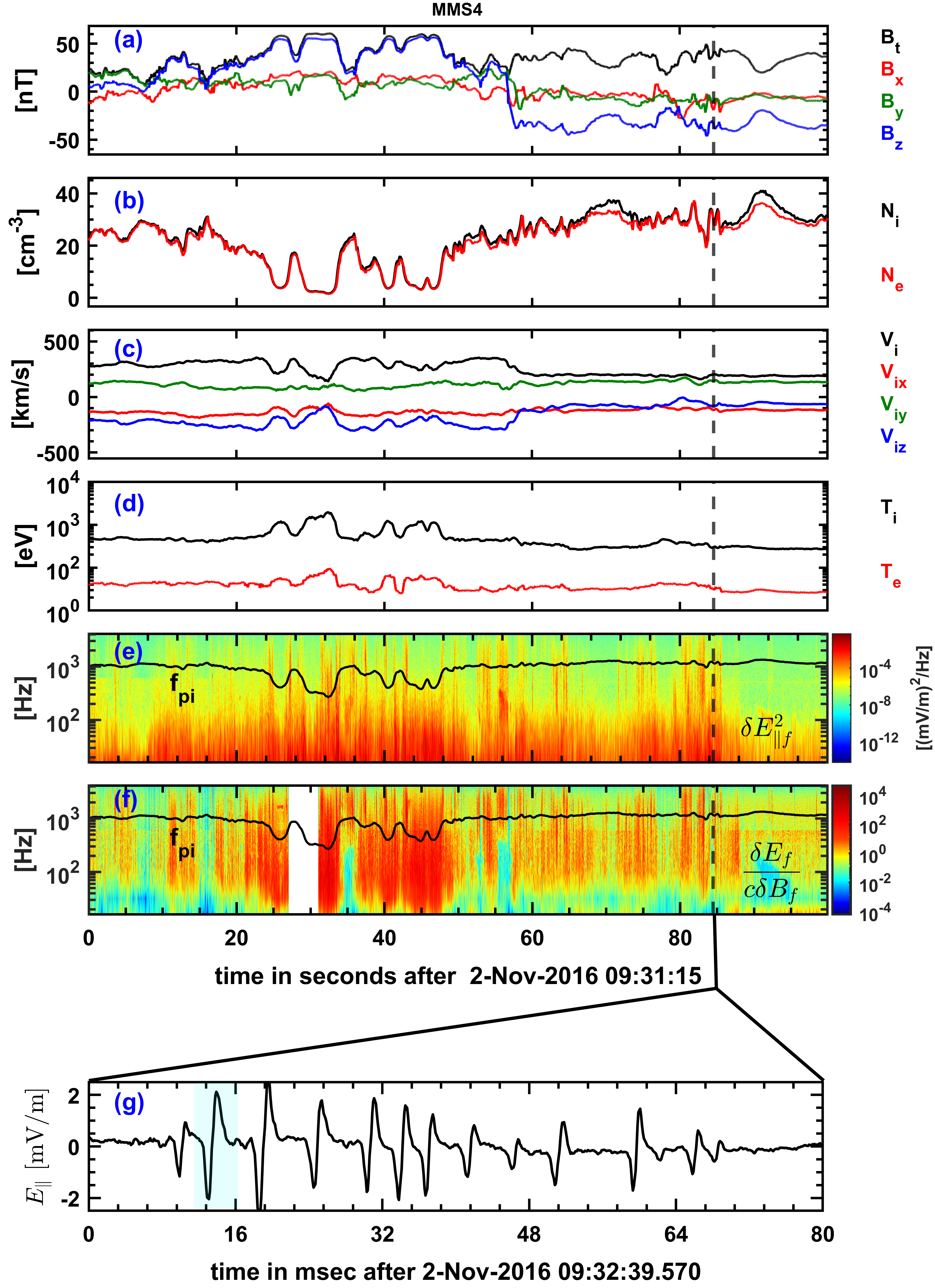}
\caption{Overview of MMS4 burst mode measurements in the Earth's magnetosheath on November 2, 2016: (a) the magnetic field, (b) electron and ion densities, (c) ion flow velocity, (d) ion and electron temperatures, (e) power spectral density of parallel electric field fluctuations computed using Fast Fourier Transform with a 100 ms sliding window (ion plasma frequency $f_{pi}$ is indicated in the panel), (f) the ratio $\delta E_{f}/c\delta B_{f}$ between total power spectral densities of electric and magnetic field fluctuations ($c$ is the speed of light), (g) the electric field measured over about $100$ ms demonstrating the train of electrostatic solitary waves. All vector quantities are in the Geocentric Solar Ecliptic (GSE) coordinates.} 
\label{Fig:IP}
\end{figure}

Figure \ref{Fig:IP} overviews \blue{about two minutes} of MMS4 observations in the Earth's magnetosheath. Note that other MMS spacecraft were located within about ten kilometers of MMS4 and provided \blue{essentially} identical overviews. Panels (a)--(c) show that the magnetic field varied between 20 and 50 nT, the plasma density was between 5 and 40 cm$^{-3}$, and plasma flow velocity was about 200--300 km/s. Panel (d) shows that ions were by a factor of ten hotter than electrons, $T_i \approx 0.3$--1 keV and $T_e \approx 20$--50 eV. Ions were essentially protons since the density of helium and oxygen ions were less than 1\% of the proton density (Supporting Materials; SM). The power spectral density of parallel electric field fluctuations in panel (e) demonstrates the presence of broadband electrostatic fluctuations up to about \blue{the} ion plasma frequency, $f_{pi}\approx 0.3$--1 kHz. The ratio between the total power spectral densities of electric and magnetic field fluctuations in panel (f) shows that $\delta E_{f}/\delta B_{f}\gg c$ above about 100 Hz ($c$ is the speed of light), which implies the broadband fluctuations are predominantly electrostatic in accordance with earlier spectral measurements \citep{Rodriguez79:jgr,Mangeney06:angeo}. A visual inspection of electric field waveforms reveals that the broadband fluctuations correspond particularly to electrostatic solitary waves exemplified in panel (g). These solitary waves with typical temporal widths of 1 ms and predominantly parallel electric fields of a few mV/m are similar to those reported previously by \cite{Kojima97:jgr} and \cite{Graham16:jgr}. Note that the train of solitary waves in Figure \ref{Fig:IP}g was previously analyzed using {\it multi-spacecraft} interferometry \citep{Holmes18:jgr}.


\begin{figure}
\centering
\includegraphics[width=1\linewidth]{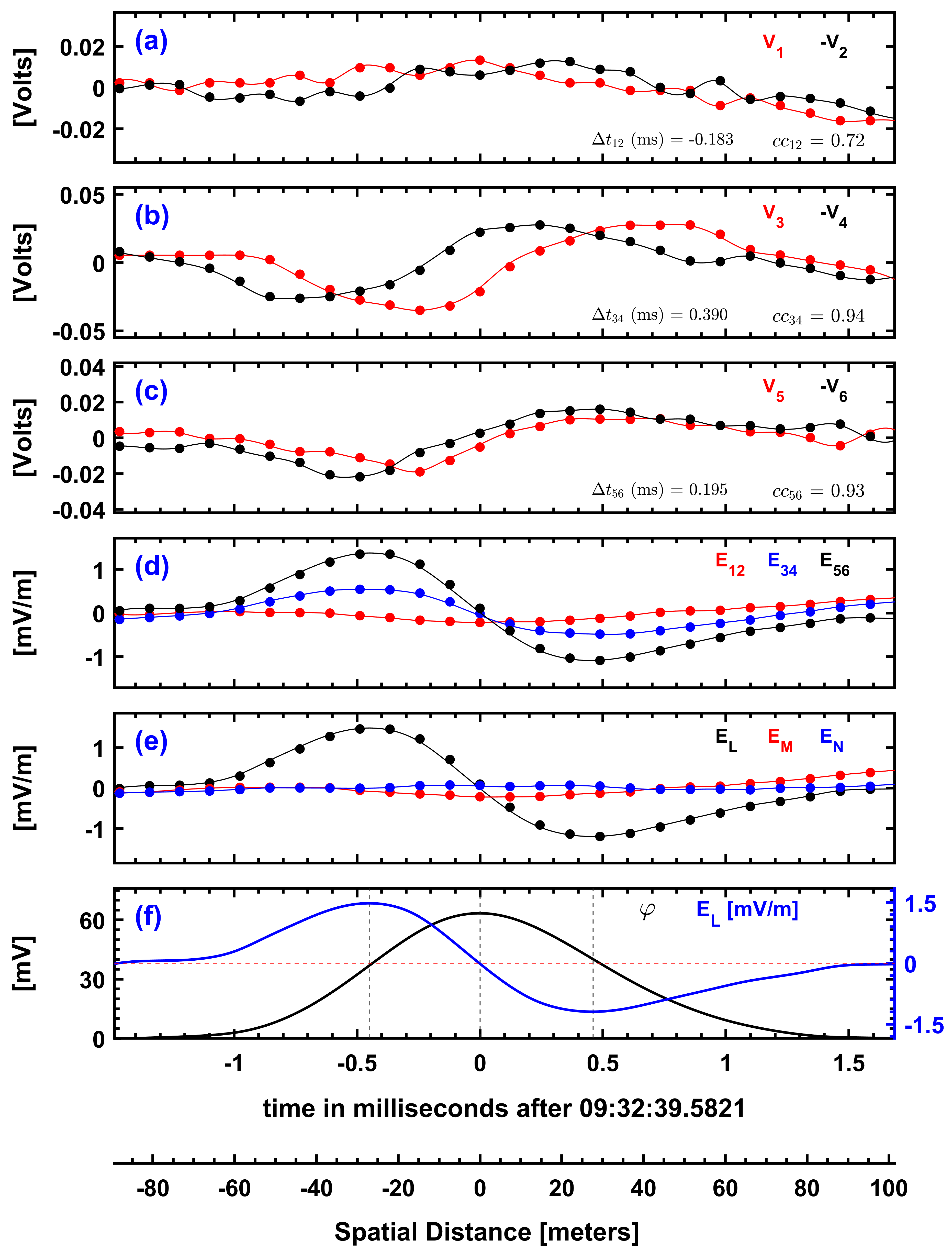}  
\caption{Interferometry analysis of the solitary wave highlighted in Figure \ref{Fig:IP}g: (a)--(c) voltage signals of spin plane ($V_1$--$V_{4}$) and axial probes ($V_{5}$ and $V_{6}$) measured with respect to spacecraft and high-pass filtered above 10 Hz to remove offsets; time delays and correlation coefficients between the voltage signals of opposing probes are indicated in the panels; (d) the electric field components in the coordinate system related to the electric field antennas: $E_{12} = 1.35\;(V_2 - V_1)/120\;{\rm m}$, $E_{34} = 1.35\;(V_4 - V_3)/120\;{\rm m}$, and $E_{56} = 1.2\;(V_6 - V_5)/29.2\;{\rm m}$; (e) electric field components in the coordinate system {\bf LMN} obtained by applying Maximum Variance Analysis \citep{Sonnerup&Scheible98} to the electric field $(E_{12},E_{34},E_{56})$ in panel (d); (f) the electrostatic potential, $\varphi = \int E_L V_s dt$, where $V_{s}=60$ km/s is the solitary wave velocity in the spacecraft frame estimated using the time delay between $V_3$\&$-V_4$. Panels (a)--(e) indicate the actual measurements at 8,192 S/s resolution by dots, while the continuous solid curves correspond to spline interpolated data. The bottom axis shows the spatial coordinate $V_s (t-t_0)$ along the electric field polarization direction ${\bf L}$, where $t_0$ corresponds to the middle between the electric field peaks.} 
\label{Fig:V-Interferometry}
\end{figure}

Figure \ref{Fig:V-Interferometry} demonstrates {\it single-spacecraft} interferometry for the solitary wave highlighted in Figure \ref{Fig:IP}g. Panels (a)--(c) present voltage signals of the spin plane and axial probes high-pass filtered above 10 Hz to remove offsets. Panel (d) presents the electric field computed using the voltage signals:  $E_{12}=1.35\cdot (V_{2}-V_{1})/2l_{12}$, $E_{34}=1.35\cdot(V_{4}-V_{3})/2l_{34}$ and $E_{56}=1.2\cdot (V_{6}-V_{5})/2l_{56}$, where $l_{12}=l_{34}=60$ m and $l_{56}=14.6$ m are spin plane and axial antenna lengths, 1.35 and 1.2 are the optimal frequency response factors \citep{Wang21:jgr}. Applying Maximum Variance Analysis \citep{Sonnerup&Scheible98} to the electric field in panel (d), we obtain a unit maximum variance vector, ${\bf L}\approx (-0.01, 0.39, 0.92)$, in the coordinate system related to the electric field antennas. Panel (e) shows that the electric field component $E_{L}$ strongly dominates components perpendicular to {\bf L}, which implies that ${\bf L}$ gives the electric field polarization direction and also the propagation direction (ambiguous by $180^{\circ}$ though) provided that the solitary wave is locally planar and one-dimensional. Voltage signals $V_{3}$ \& $-V_{4}$ and $V_{5}$ \& $-V_{6}$ exhibit the highest cross-correlation coefficients and the largest time delays, $\Delta t_{34} \approx  0.39$ ms and $\Delta t_{56}\approx -0.18$ ms. \blue{The voltage signals were up-sampled by a factor of ten to increase the accuracy of the time delays. The up-sampling procedure presumes the voltage signals are sufficiently smooth, which is justified by the smoothness of the electric field measured at a higher resolution of 65536 S/s (not shown here).} Assuming local planar and one-dimensional solitary wave, we use time delays between the pairs of opposing probes to obtain independent estimates of the solitary wave velocity in the spacecraft frame, $V_{s}={\bf L}_{ij}l_{ij}/\Delta t_{ij}$ \citep{Vasko20:front,Wang21:jgr}. Using the time delays between $V_{3}$\&$-V_{4}$ and $V_{5}$\&$-V_{6}$, we obtain independent velocity estimates of 60 and 68 km/s, whose consistency strongly indicates that the solitary wave was indeed locally planar and one-dimensional. The positive value of the estimated velocity indicates that the solitary wave propagates parallel to the chosen polarization direction ${\bf L}$ and almost anti-parallel (within about $10^{\circ}$) to \blue{the} local magnetic field. The use of the time delay between $V_{1}$\&$-V_{2}$ provides the speed estimate of only a few km/s, but these voltage signals were poorly correlated, and the solitary wave was barely propagating along this antenna (${\bf L}_{12}\approx -0.01$), which makes the speed estimate highly uncertain \citep{Wang21:jgr}. The revealed solitary wave velocity of about 60 km/s is consistent with the results of electric field interferometry (SM) which is another technique for estimating wave velocities \citep[e.g.,][]{Graham16:jgr,Steinvall:jgr22}. Using solitary wave velocity $V_{s}=60$ km/s, we compute the electrostatic potential $\varphi=\int E_{L}\;V_{s}\;dt$ and also translate temporal solitary wave profiles into spatial ones. Panel (f) shows that the solitary wave has a {\it positive} peak of the electrostatic potential, $\varphi_0\approx 60$ mV, and \blue{the parallel} half-width \blue{that is the distance} between electric field peaks of $l\approx 29$ m. In units of local electron temperature and Debye length, we have $e\varphi_0/T_e\approx 2\cdot 10^{-3}$ and $l/\lambda_{D}\approx 3.5$.

\begin{figure}
\centering
\includegraphics[width=\linewidth]{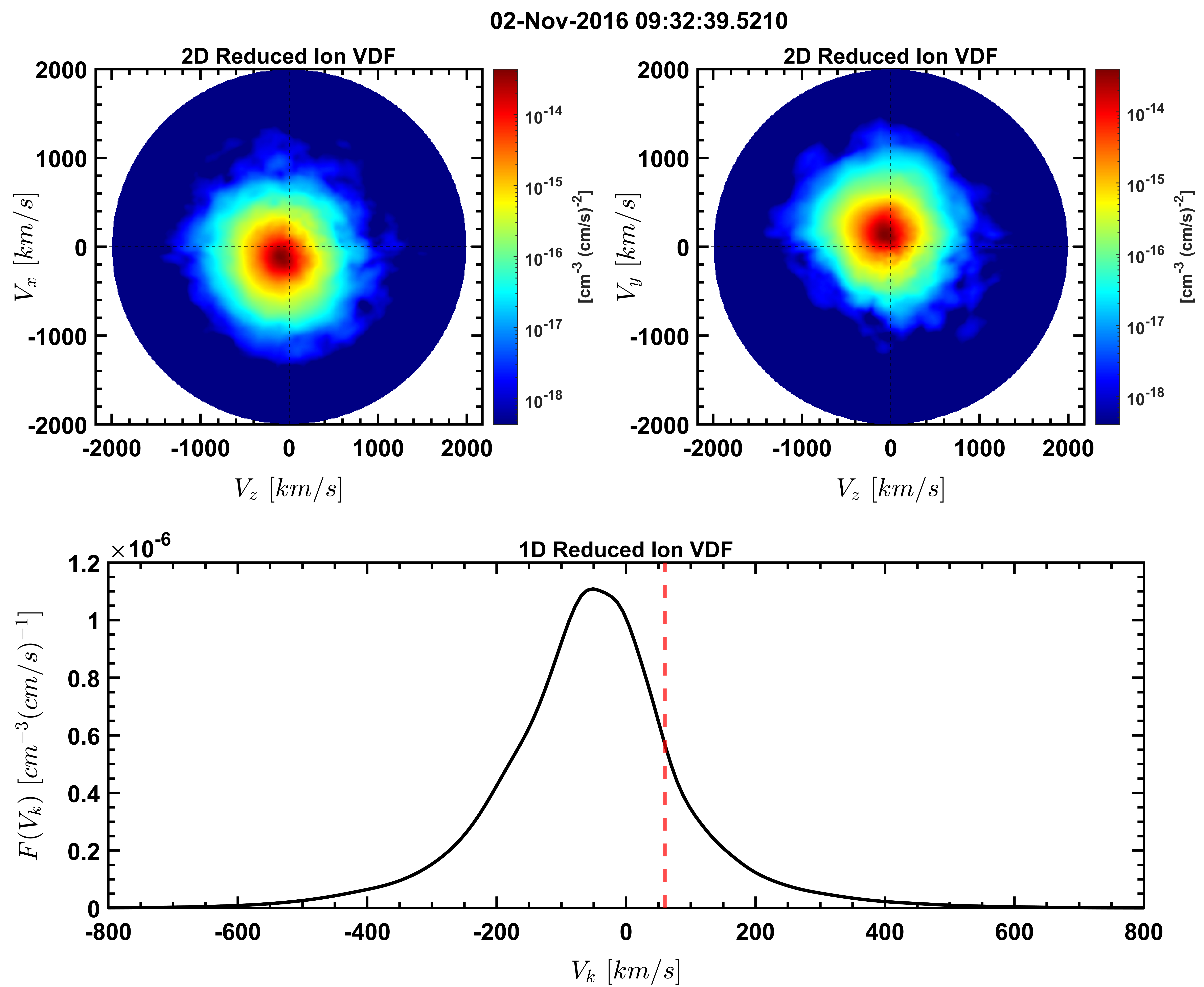}
\caption{Ion velocity distribution functions in the spacecraft frame collected over 150 ms around the solitary wave shown in Figure \ref{Fig:V-Interferometry}; note that the solitary wave propagates almost along the axial antenna or, equivalently, almost along the $z-$axis in the GSE. The top panels present 2D reduced distributions, $\int {\rm VDF}({\bf V})\;{\rm d}V_y$ and $\int {\rm VDF}({\bf V})\;{\rm d}V_x$, where the GSE coordinates are used. The bottom panel presents a 1D distribution obtained by reducing the 3D distribution over two velocities perpendicular to the solitary wave propagation direction, ${\bf k}_{\rm GSE} \approx (0.32, 0.23, 0.92)$. The vertical red line in the bottom panel corresponds to the solitary wave velocity in the spacecraft frame.} 
\label{Fig:ion_VDF}
\end{figure}

\begin{figure}
\centering
\includegraphics[width=\linewidth]{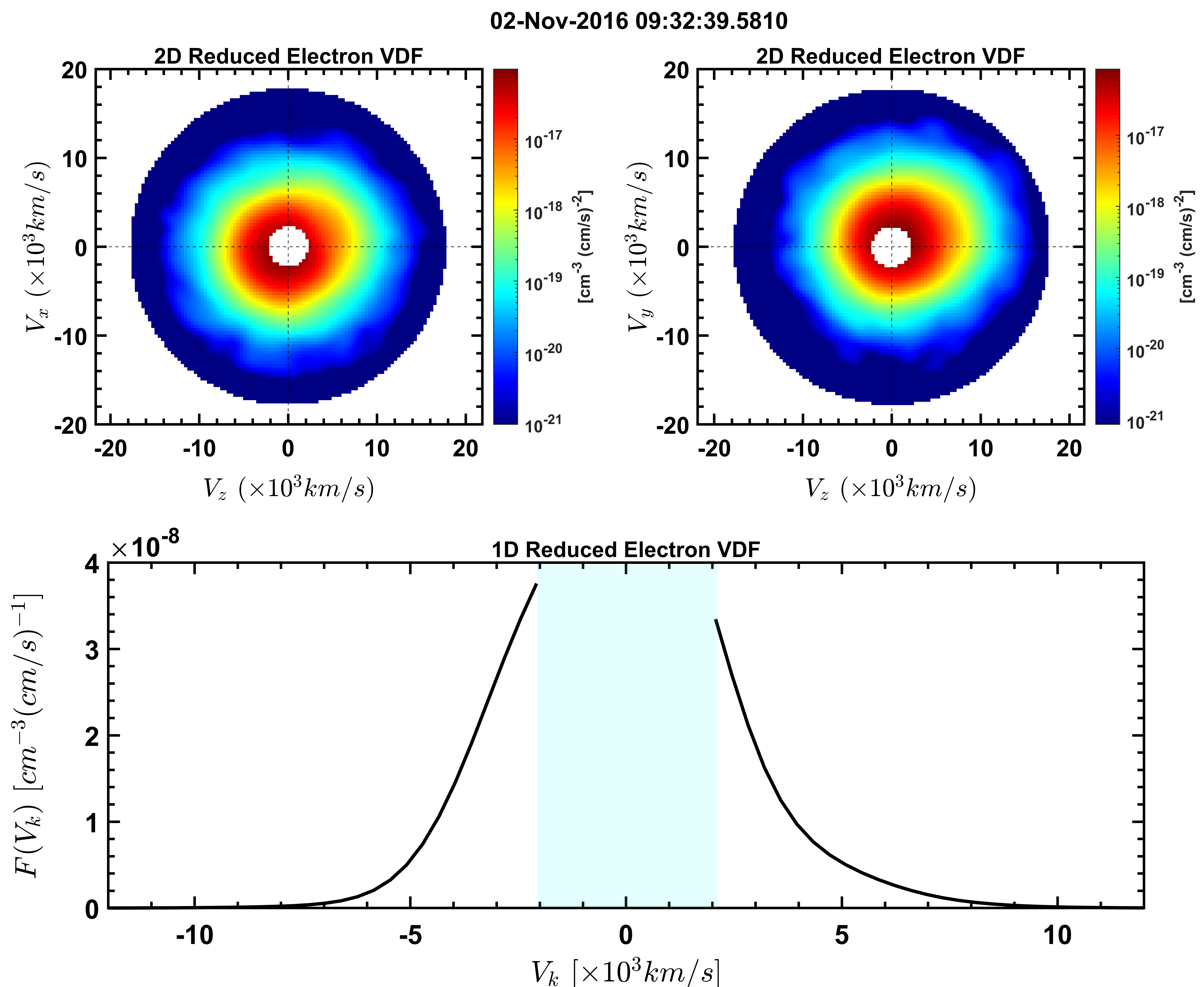}
\caption{Electron velocity distribution functions in the spacecraft frame collected over 30 ms around the solitary wave shown in Figure \ref{Fig:V-Interferometry}. The format of the figure is identical to that of Figure \ref{Fig:ion_VDF}. The electron distribution below about 20 eV is contaminated by photoelectrons and was cut out in both the upper and bottom panels.} 
\label{Fig:electron_VDF}
\end{figure}

Figures \ref{Fig:ion_VDF} and \ref{Fig:electron_VDF} present ion and electron velocity distribution functions (VDF) measured aboard MMS4 around the solitary wave highlighted in Figure \ref{Fig:IP}g. The upper panels demonstrate 2D reduced distributions, $\int {\rm VDF}({\bf V})\;{\rm d}V_{y}$ and $\int {\rm VDF}({\bf V})\;{\rm d}V_{x}$, while the bottom panels present 1D distributions $F(V_{k})$ computed by reducing 3D VDFs over two velocities perpendicular to the solitary wave propagation direction, ${\bf k}={\bf L}\cdot {\rm sign}(V_{s})$; the latter \blue{is} basically along the axial antenna (Figure \ref{Fig:V-Interferometry}d) that is essentially along the $z-$axis, ${\bf k}_{\rm GSE} \approx (0.32, 0.23, 0.92)$. \blue{The spacecraft potential} was positive and around 3 V (not shown), but the electron fluxes below about 20 eV were contaminated by \blue{secondary electrons and photoelectrons}. The corresponding part of the electron distribution is cut out, while the distribution above about 20 eV does not exhibit any signatures of electron beams. \blue{The ions have} a quasi-Maxwellian distribution without any signature of beams as well. The drift between ions and electrons determined by computing the local current density using the curlometer technique \citep[e.g.,][]{Chanteur98:issi} was within one hundred km/s (SM). The solitary wave velocity is clearly shifted with respect to the peak of the 1D reduced ion distribution. In accordance with that the plasma frame velocity of the solitary wave \blue{is} $V_{s}^*=V_{s}-{\bf V}_{i}\cdot {\bf L}\approx 138$ km/s, where both ion flow velocity ${\bf V}_i$ and electric field polarization direction ${\bf L}$ \blue{are} in the GSE coordinates. The plasma frame speed is well below local electron thermal speed but comparable with ion thermal speed: $V_{s}^*\approx 0.04\;V_{Te}$ and $V_s^*\approx 0.6\;V_{Ti}$, where $V_{Te}=(2T_{e}/m_{e})^{1/2}\approx 3,300$ km/s and $V_{Ti}=(2T_{i}/m_{p})^{1/2}\approx 240$ km/s.

Similar single-spacecraft analysis of other solitary waves shown in Figure \ref{Fig:IP}g revealed all of them have {\it positive polarity}, \blue{parallel} half-widths of about 30 m that is a few Debye lengths and spacecraft frame velocities of about 60 km/s. Note that no distortion of spin plane electric field components was observed for the solitary waves in Figure \ref{Fig:IP}g, even though their spatial half-widths were less than the separation between spin plane probes \citep[see the analysis of such distortions in][]{Wang21:jgr}. The distortions expected for these, as well as all other solitary waves discussed in this study, were absent, because the solitary waves propagated almost along the axial antenna \blue{that is} at large angles to the spin plane \citep[see details in][]{Wang21:jgr}. The revealed solitary wave parameters are drastically different from those ({\it negative polarity}, widths of about 1 km or a few hundred Debye lengths and speeds of \blue{the order of} 1,000 km/s) obtained via multi-spacecraft interferometry \citep{Holmes18:jgr}. We believe the multi-spacecraft interferometry failed, because seemingly correlated waveforms observed aboard different spacecraft actually correspond to different solitary waves. This implies that the spatial extent of the solitary waves in the plane perpendicular to the local magnetic field was smaller than the spatial separation between MMS spacecraft of about 10 km. Note that the perpendicular extent of even 1 km would not contradict local planarity of the solitary waves, since their parallel spatial \blue{half-widths} were less than 100 m. All the solitary waves were associated with quasi-Maxwellian ion distributions and had plasma frame velocities comparable with ion thermal speed and much smaller than electron thermal speed. In the next section, we present the results of a similar analysis for a statistically representative dataset of solitary waves collected in the considered interval.




\section{Statistical results\label{sec3}}

We visually inspected electric field waveforms measured aboard four MMS spacecraft over the interval shown in Figure \ref{Fig:IP} and collected more than 2,000 solitary waves with bipolar profiles and peak-to-peak temporal widths exceeding $4\times 0.12$ ms$\approx 0.5$ ms. The latter criterion ensures at least four points of electric field measurements between the electric field peaks \blue{and, hence, a decent resolution of solitary wave waveform}. For each solitary wave we computed (1) correlationthe  coefficients and time delays between the voltage signals of opposing probes and (2) maximum variance vector ${\bf L}$ along with intermediate and minimum variance vectors (${\bf M}$ and ${\bf N}$) and corresponding variances $\lambda_{\rm max}$, $\lambda_{\rm int}$ and $\lambda_{\rm min}$. The interferometry could be performed only for \blue{645} solitary waves with (1) at least one pair of the voltage signals having a correlation coefficient higher than 0.85 and corresponding time delay larger than 0.06 ms, and (2) $\lambda_{\rm max}/(\lambda_{\rm int}^2+\lambda_{\rm min}^2)^{1/2}>5$ that is equivalent to $E_{L}\gg E_{M}, E_{N}$ and \blue{strictly necessary to ensure local} solitary wave planarity. \blue{The latter condition ensures that the solitary waves have perpendicular spatial extent of at least about five times larger than their parallel width, which is sufficient to consider the solitary waves planar on the scale of spatial separation between voltage-sensitive probes (Section \ref{sec5}).} The collected solitary waves have peak-to-peak temporal widths and electric field amplitudes, respectively, around 1 ms and 1 mV/m (SM). For each solitary wave, we computed velocities in spacecraft and plasma frames, the amplitude of the electrostatic potential, spatial half-width, and associated particle distribution functions. \blue{We revealed 15} solitary waves of negative polarity, but those were excluded from \blue{the present analysis} (their speed and spatial width were similar to those of the solitary waves of positive polarity) and \blue{in what follows we focus on the 630 solitary waves of positive polarity}. 

\begin{figure}
\centering
\includegraphics[width=\linewidth]{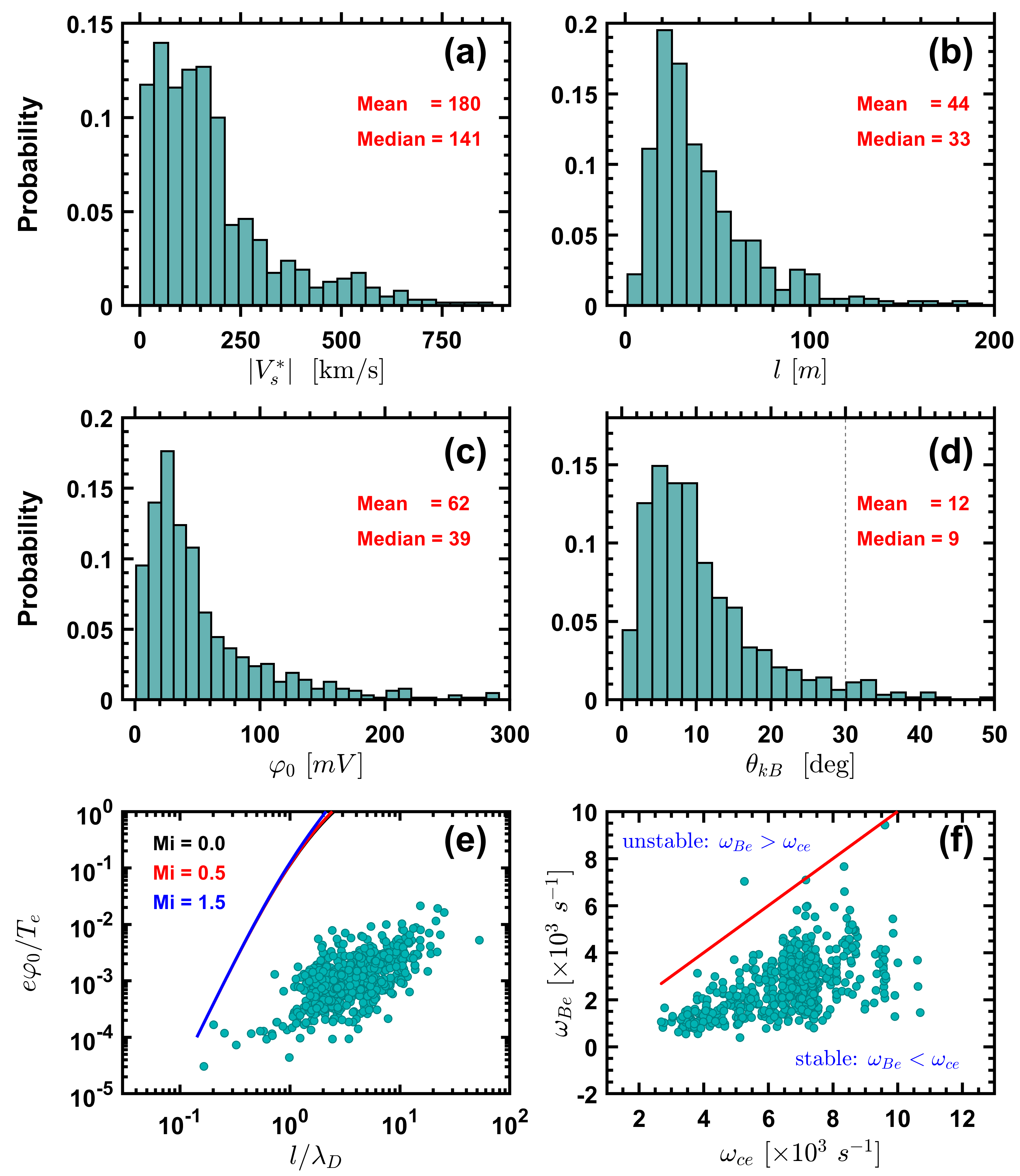}
\caption{Panels (a)--(d) present distributions of various properties of the 630 solitary waves \blue{of positive polarity}: (a) plasma frame speeds $|V_{s}^{*}|$, (b) \blue{parallel} spatial half-widths $l$, (c) peak values $\varphi_0$ of the electrostatic potential, (d) angles $\theta_{kB}$ between local magnetic field and solitary wave propagation direction or, equivalently, electric field polarization direction. Panel (e) shows normalized solitary wave amplitudes $e\varphi_0/T_{e}$ versus their normalized half-widths $l/\lambda_{D}$. The curves in panel (e) correspond to theoretical inequality (\ref{eq:B1}) between $l/\lambda_{D}$ and $e\varphi_0/T_{e}$ that should be satisfied by electron holes (Appendix \ref{B} Width-amplitude relation); the curves are shown for several typical solitary wave Mach numbers, $M_{i}=|V_{s}^{*}|/V_{Ti}$, and electron-to-ion temperature ratio of $T_{e}/T_{i}=0.1$. Since all the solitary waves satisfy inequality (\ref{eq:B1}), they can be, in principle, electron holes. Panel (d) presents electron cyclotron frequency $\omega_{ce}$ versus $\omega_{Be}=(e\varphi_0/m_{e}l^2)^{1/2}$, where the latter is the bounce frequency of electrons trapped within positive potential with peak value $\varphi_0$ and spatial half-width $l$. Electron holes stable to the transverse instability should have $\omega_{Be}\lesssim \omega_{ce}$ \citep{Muschietti00,Hutchinson18:prl}.} 
\label{Fig:histogram}
\end{figure}

Figure \ref{Fig:histogram} presents the properties of \blue{the 630} solitary waves estimated by single-spacecraft interferometry. Panels (a) and (b) present statistical distributions of spacecraft frame speeds and \blue{parallel} half-widths of the solitary waves. The solitary waves have plasma frame speeds typically within 300 km/s and \blue{parallel} half-widths typically within 100 m; the corresponding median values are 130 km/s and 30 m. Panel (c) demonstrates that all the solitary waves have {\it positive polarity} and amplitudes from 10 to 200 mV with a median value of about 40 mV. Panel (d) shows that all the solitary waves propagate quasi-parallel to local magnetic field lines, typically within about 20$^{\circ}$. The normalized solitary wave amplitudes $e\varphi_0/T_e$ and half-widths $l/\lambda_{D}$ in panel (e) demonstrate that the solitary waves have typical amplitudes within 0.01-1\% of local electron temperature and half-widths between about 1 and 10 Debye lengths. Note a moderate positive correlation between $e\varphi_0/T_{e}$ and $l/\lambda_{D}$; solitary waves with larger amplitudes tend to have larger spatial widths.




\begin{figure}
\centering
\includegraphics[width=\linewidth]{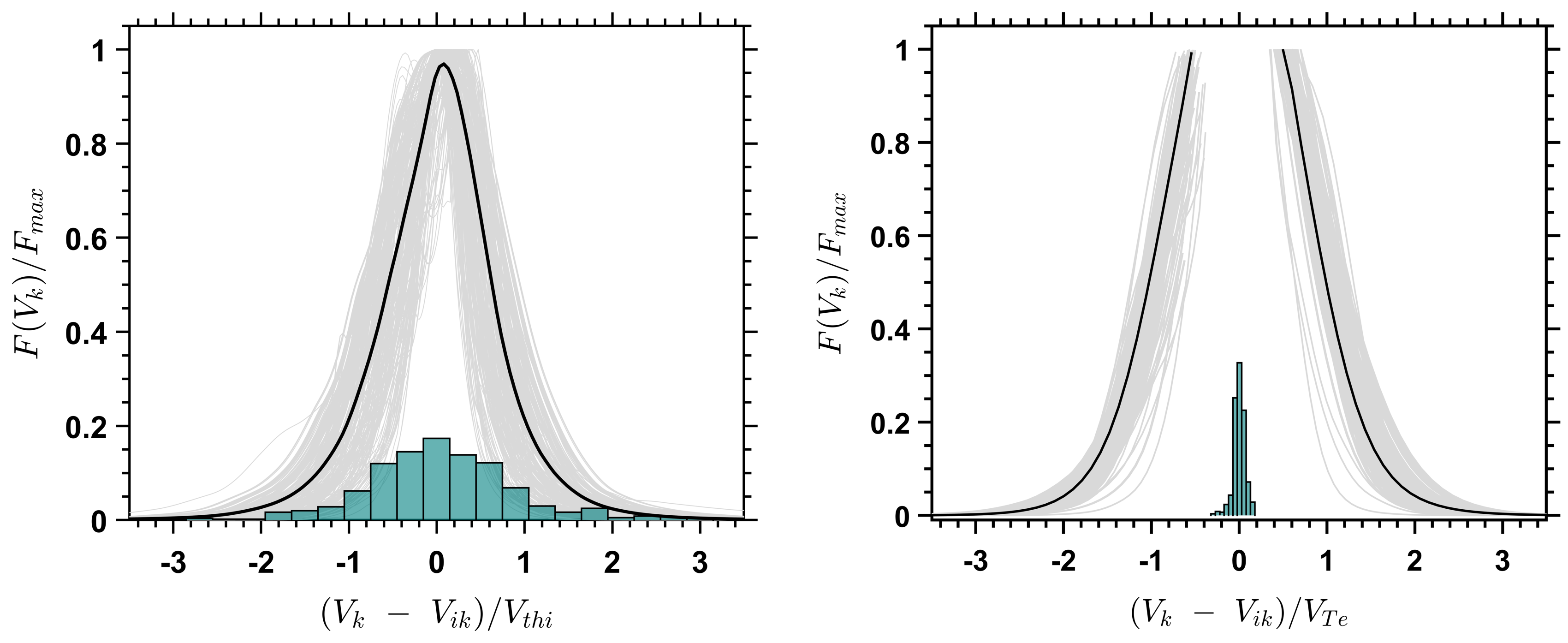} 
\caption{One-dimensional reduced velocity distributions $F(V_{k})$ of ions and electrons associated with the 630 solitary waves \blue{of positive polarity}; each velocity distribution was computed by reducing corresponding three-dimensional distribution over two velocities perpendicular to solitary wave propagation direction ${\bf k}$. Each reduced distribution $F(V_{k})$ was normalized to its peak value and translated into the plasma frame with velocities normalized to electron or ion thermal speed. In both panels, $V_{ik}$ is the ion flow velocity along the solitary wave propagation direction, $V_{Te}$ and $V_{Ti}$ are, respectively, electron and ion thermal speeds. The black curves present ion and electron distributions obtained by averaging velocity distributions corresponding to individual solitary waves (Section \ref{sec3}). The individual and averaged electron distributions are missing at velocities smaller than about 0.2$V_{Te}$ because of contamination by photoelectrons \blue{and secondary electrons} at energies below about 20 eV. The histograms present distributions of solitary wave velocities in the plasma frame normalized to ion and electron thermal speeds.} 
\label{Fig:Reduced_VDF}
\end{figure}

Figure \ref{Fig:Reduced_VDF} presents 1D ion and electron distributions associated with each of the 630 solitary waves. These 1D reduced distributions were computed similarly to those in Figures \ref{Fig:ion_VDF} and \ref{Fig:electron_VDF}, but presented in a different way. First, each 1D distribution $F(V_{k})$ was normalized to its peak value and considered in the plasma frame with velocities $V_{k}-V_{ik}$ normalized to the corresponding thermal speed, where $V_{ik}$ is the ion flow velocity along solitary wave propagation direction in the spacecraft frame. Second, ion distributions with the peak at $V_{k}<V_{ik}$ were reflected to have the peak at $V_{k}>V_{ik}$. If the ion distribution had to be reflected, the corresponding electron distribution and solitary wave velocity $V_{s}^{*}$ were reflected too. Both normalization and reflection were necessary to compute averaged particle distributions and treat equally particle distributions corresponding to individual solitary waves in the averaging process. The averaged ion distribution presented in panel (a) is clearly quasi-Maxwellian and exhibits no signatures of beams. The averaged electron distribution also exhibits no signatures of beams at velocities larger than about $0.2V_{Te}$. At smaller velocities, the electron distributions were contaminated by photoelectrons and \blue{secondary electrons}. The histograms in panels (a) and (b) present normalized plasma frame velocities of the solitary waves. The solitary waves have plasma frame speeds well below electron thermal speed but comparable with typical velocities of the bulk of ions. For more than 75\% of the solitary waves we have $|V_{s}^{*}|<0.05\;V_{Te}$ and $|V_{s}^{*}|\lesssim V_{Ti}$.


\begin{figure}
\centering
\includegraphics[width=0.8\linewidth]{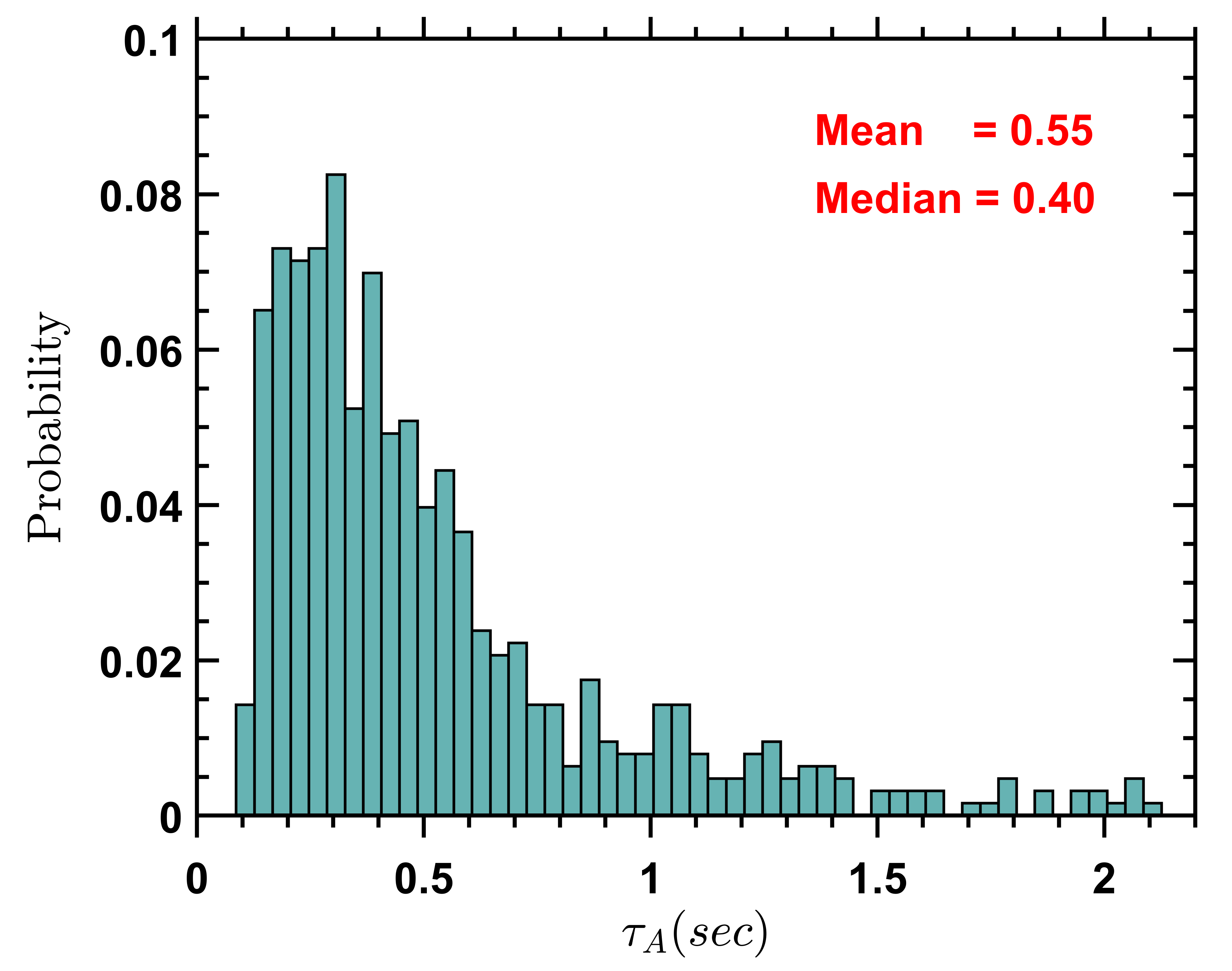}
\caption{The distribution of acceleration time scales (e-folding times) of the 630 slow electron holes that were computed using Eq. (\ref{eq:tau_A}). Note that the acceleration of a slow electron hole occurs due to interaction with ions \citep{Hutchinson21:pre,Hutchinson23:phpl}.} 
\label{Fig:tau_A}
\end{figure}

\section{Theoretical interpretation and estimates}

One of the types of electrostatic solitary waves in plasma are solitons, \blue{structures} that can be described within fluid approximation \citep{Sagdeev66,Lotko83,Lakhina09:jgr}. The solitary waves discussed in the previous section are highly unlikely to be any type of solitons. First, the width and amplitude of small-amplitude solitons are strictly dependent as $e\varphi_{0}/T_{e}\propto (\lambda_{D}/l)^{2}$ \citep[][]{Sagdeev66,Lotko83}, while the observed solitary waves tend to have larger amplitudes for larger spatial widths (Figure \ref{Fig:histogram}e). Second, theoretical analysis shows that the Poisson equation with electron and ion response computed from fluid equations does not have solutions that would correspond to the observed solitary waves (Appendix \ref{A} Fluid models). Note that in the latter theoretical analysis, we assumed the electron distribution below about 20 eV to be Maxwellian (the electron and ion densities match well after this extrapolation), while the electron distribution at these low energies is not known. \blue{We believe that small-amplitude solitary waves (if corresponding solutions existed) would display soliton-like scaling, $e\varphi_{0}/T_{e}\propto (\lambda_{D}/l)^{2}$, even in the case of more complex electron distributions at the low energies and this scaling would not be corroborated by observations}. The only other interpretation is that the solitary waves are {\it electron holes} that is purely kinetic structures, whose existence is due to a deficit of electrons trapped within their positive potential \citep{Roberts67,Turikov84,Schamel86,Mozer18:prl}.

The width and amplitude of electron holes must satisfy some inequality, ensuring the positiveness of the distribution function of trapped electrons \citep{Turikov84,Chen04:pre,Chen05:jgr,Goldman07:prl}. We generalized the inequality previously derived by \cite{Chen05:jgr} for the case of electron holes not steady in the plasma frame (Appendix \ref{B} Width-amplitude relation). The corresponding inequality (\ref{eq:B1}) states that $l/\lambda_{D} \geq S(e\varphi_0/T_{e},M_{i},T_{e}/T_{i})$ and depends on ion sound Mach number $M_{i}=|V_{s}^*|/V_{Ti}$ and electron to ion temperature ratio $T_{e}/T_{i}$. Theoretical curves $l/\lambda_{D}=S(e\varphi_0/T_{e}, M_{i}, T_{e}/T_{i})$ corresponding to the typical temperature ratio of $T_{e}/T_{i}=0.1$ and Mach numbers of $M_{i}=0.5, 1$ and 1.5 all show that the observed solitary waves satisfy inequality (\ref{eq:B1}). Based on this analysis, we interpret the solitary waves in terms of {\it electron holes}. Note that stable existence of electron holes is possible only when the bounce frequency of trapped electrons $\omega_{Be}=(e\varphi_0/m_{e}l^2)^{1/2}$ is below local electron cyclotron frequency $\omega_{ce}=eB/m_{e}c$ \citep{Muschietti00,Hutchinson18:prl}. Figure \ref{Fig:histogram}f shows that larger $\omega_{Be}$ occur at larger $\omega_{ce}$, but the former is statistically below the latter. Thus, the observed solitary waves are indeed stable to the transverse instability, which reinforces the interpretation of the solitary waves in terms of electron holes.

The very existence of electron holes, whose plasma frame speed is comparable with ion thermal speed, is actually surprising. According to numerical simulations and theory, in plasma with quasi-Maxwellian ions, an initially slow electron hole will accelerate to velocities much larger than ion thermal speed due to interaction with ions \citep{Eliasson&Shukla04:prl,Hutchinson21:pre,Hutchinson23:phpl}. Theory predicts that the velocity of a slow electron hole should grow exponentially, and the corresponding e-folding time can be estimated as follows 
\begin{eqnarray}
    \tau_{A}\approx \omega_{pe}^{-1} \left(1+T_{i}/T_{e}\right)^{1/2}\; (e\varphi_0/T_{i})^{-1}\;l/\lambda_{D},
    \label{eq:tau_A}
\end{eqnarray}
where $\omega_{pe}$ is electron plasma frequency \citep[see \blue{Eq. (22)} in][]{Hutchinson23:phpl}. For the observed electron holes the e-folding time $\tau_{A}$ is from a fraction to a few seconds, and the median value is around 0.5s (Figure \ref{Fig:tau_A}).

\section{Discussion and Conclusion \label{sec5}}

In this paper, we presented a statistical analysis of electrostatic solitary waves observed over \blue{about two minutes} in the Earth's magnetosheath. Similar solitary waves were previously reported in the Earth's magnetosheath \citep{Kojima97:jgr,pickett03:npg,Pickett05}, but the latter studies could not reveal their speed, spatial width and amplitude of the electrostatic potential. \cite{Graham16:jgr} and \cite{Holmes18:jgr} have recently analyzed a few tens of solitary waves in the Earth's magnetosheath but reported drastically different solitary wave properties. The inconsistency between the previous studies and limited information about solitary waves in the Earth's magnetosheath in general have motivated the present study.

We carried out single-spacecraft analysis for \blue{645 solitary waves} and \blue{found all of them except 15} to be structures of {\it positive polarity} with \blue{parallel} half-widths of 10--100 m and amplitudes of 10--200 mV, propagating quasi-parallel to a local magnetic field with typical plasma frame velocities of the order of 100 km/s. These \blue{parallel} half-widths are between about 1 and 10 Debye lengths, while the amplitudes are within 0.01--1\% of local electron temperature. \blue{All these estimates would not substantially change if we used thresholds higher than 0.85 and 0.06 ms for respectively correlation coefficients and time delays between signals of voltage-sensitive probes.} The revealed properties are drastically different from those reported for \blue{the same solitary waves} by \cite{Holmes18:jgr}, \blue{who used multi-spacecraft interferometry (assuming a solitary wave is sequentially observed aboard four MMS spacecraft) to estimate their velocities}. The multi-spacecraft interferometry resulted in order of magnitude higher velocities and {\it negative polarity} of the electrostatic potential. We believe multi-spacecraft interferometry fails because \blue{the solitary waves} have \blue{perpendicular spatial extent} well below MMS spacecraft separation of 10 km, and electric field signals observed aboard different spacecraft do not correspond to the same solitary wave. \blue{Note that the considered solitary waves could be considered planar on the scale of spatial separation between voltage-sensitive probes since the perpendicular extent was required to be at least five times larger than the parallel width and, hence, exceeded about 100 m.} We demonstrated that the solitary waves must be {\it electron holes}, that is purely kinetic structures and, more specifically, {\it slow electron holes}, since their plasma frame velocity $V_{s}^{*}$ is comparable with ion thermal speed and well below electron thermal speed, $|V_{s}^{*}|\lesssim V_{Ti}$ and  $|V_{s}^{*}|<0.05\; V_{Te}$. Note that solitary wave properties revealed in our analysis are similar to those of a few tens of solitary waves reported in the Earth's magnetosheath by \cite{Graham16:jgr}.

Slow electron holes are expected to efficiently interact with ions, which generally prevents them from remaining slow and accelerates them to velocities much higher than ion thermal speed \citep{Eliasson&Shukla04:prl,Hutchinson21:pre,Hutchinson23:phpl}. This acceleration process can only be avoided, when the distribution function of background ions has a sufficiently deep local minimum, and the electron hole velocity is around that local minimum \citep{Hutchinson21:pre}. \cite{Kamaletdinov21:prl} have recently shown that slow electron holes in the Earth's plasma sheet are indeed associated with double-hump ion distribution functions and have their velocity around the local minimum. In contrast, the slow electron holes in the Earth's magnetosheath are associated with quasi-Maxwellian ion distributions and are expected to accelerate. We estimated the acceleration time scale to be on the order of one second. Since all the observed electron holes are slow, their lifetime is expected not to significantly exceed the acceleration time scale. This upper estimate on the lifetime does not contradict the lifetime of 1,000\;$\omega_{pe}^{-1}$ observed in numerical simulations \citep[][]{Oppenheim99:prl,Oppenheim01:grl,Goldman99:grl}, which is of the order of 10 ms in the Earth's magnetosheath.

Similar slow electron holes were previously observed in the Earth's bow shock \citep{Vasko20:front,Kamaletdinov22:eh}, magnetopause \citep{Graham15,Steinvall19:grl} and plasma sheet \citep{Norgren15:exp,Lotekar20:jgr,Kamaletdinov21:prl}. The origin of slow electron holes observed in space plasma is puzzling. The slow electron holes in the Earth's magnetosheath could be, in principle, produced by Buneman or ion-acoustic instability \citep{Drake:sci2003,Buchner&Elkina06,Che10:grl}, but for typical electron to ion temperature ratio of $T_{e}/T_{i}\approx 0.1$ both instabilities would require the drift velocity between electrons and ions comparable to electron thermal speed \citep[e.g.,][]{Mikhailovskii75,Petkaki03:jgr}. In contrast, the observed drift velocity does not exceed a few hundred km/s (SM), which is negligible compared to electron thermal speed. The slow electron holes can be, in principle, produced locally by low-energy electron beams, whose distribution function cannot be observed because of the contamination by photoelectrons and \blue{secondary electrons} below about 20 eV. The alternative is that the slow electron holes originate from a distant generation region; the lifetime of a second would imply the solitary waves were observed within ten to hundred kilometers of their generation region.


In conclusion, it is appropriate to make a few comments. First, the considered solitary waves have peak-to-peak temporal widths of about 1 ms and, thus, are similar to solitary waves reported previously by \cite{Kojima97:jgr}. Solitary waves with temporal widths less than 0.1 ms are also abundant in the Earth's magnetosheath \citep{pickett03:npg,Pickett05}, but single-spacecraft interferometry is not feasible for them because voltage signals aboard Magnetospheric Multiscale are only provided at 8,192 S/s resolution. Second, while the interpretation of the solitary waves in terms of electron holes is the key result of this paper, even more important is that these structures are slow and can resonantly interact with both electrons and the bulk of ions. This implies that the solitary waves can potentially facilitate energy exchange between ions and electrons. Finally, \blue{about 2\% of the considered solitary waves (\blue{15 out of 645}) had negative polarity of the electrostatic potential and were excluded to focus on the analysis of the 630 slow electron holes}. The occurrence \blue{of solitary waves of negative polarity} was negligible in the considered interval, but may be different in the Earth's magnetosheath in general.

\appendix
%

\section{Fluid models\label{A}}

In this section, we demonstrate that the observed solitary waves {\it cannot} be described within fluid approximation with Maxwellian electrons. The electrostatic potential $\varphi$ of a one-dimensional solitary wave must satisfy the following Poisson equation
\begin{eqnarray}
\partial^{2}\varphi/\partial x^2=4\pi e(n_{e}-n_{i}),
\label{eq:A0}
\end{eqnarray}
where $n_e$ and $n_i$ are electron and ion densities. Since the observed solitary waves have plasma frame speeds well below the electron thermal speed, the electron response in fluid approximation (resonant electrons neglected) is accurately described by the Boltzmann distribution \citep{Sagdeev66,Gurevich68:jetp}. In the case of a Maxwellian distribution of background electrons we have $n_{e}/n_0=\exp(e\varphi/T_{e})=1 +e\varphi/T_e+(e\varphi/T_e)^2/2+...$, where the asymptotic expansion is valid in a small-amplitude limit, $e\varphi/T_{e}\ll 1$. The computation of the ion response is more involved, since the solitary wave speeds are smaller than or comparable with the ion thermal speed. The ion response could be computed using the ion momentum equation but would depend on {\it ad-hoc} ion polytrope index. The alternative is to compute the ion response using the kinetic equation for the bulk of ions that are not resonant with a solitary wave \citep{Tran79,Lotko83}. This approach is equivalent to fluid approximation, but automatically accounts for a proper polytrope index. For small-amplitude solitary waves, the distribution function of non-resonant ions can be expanded as follows 
\begin{eqnarray}
F_{i}=F_0(V)+\frac{e\varphi}{m_{i}(V-V_s^*)}\frac{dF_0}{dV}+\frac{e^2\varphi^2}{2m_{i}^2}\frac{1}{V-V_s^*}\frac{d}{dV}\left[\frac{1}{V-V_s^*}\frac{dF_0}{dV}\right]+...,
\label{eq:A1}
\end{eqnarray}
where $F_0(V)$ is the unperturbed ion distribution function, $V_s^*$ is the solitary wave velocity in the plasma frame and $|V-V_{s}^*|\gtrsim (2\varphi/m_{i})^{1/2}$ for non-resonant ions \citep{Tran79,Lotko83}. In the case of a Maxwellian distribution of background ions, the corresponding ion density is expanded as follows
\begin{eqnarray}
    n_{i}/n_0=1+\left(e\varphi/2T_{i}\right)\;Z^{'}_{r}(M_i)+\left(e\varphi/4T_{i}\right)^2\;Z^{'''}_{r}(M_i)+...,
    \label{eq:A2}
\end{eqnarray}
where $M_i=V_s^*\;(2T_{i}/m_i)^{-1/2}$ is the Mach number and $Z_{r}(\xi)=-2\;{\rm exp}(-\xi^2)\int _0^{\xi} {\rm exp}(t^2)\;dt$ is the real part of the Fried-Conte plasma dispersion function \citep{Tran79,Lotko83}. Normalizing the electrostatic potential to $T_{e}/e$ and the spatial scale to electron Debye length $\lambda_{D}=(T_{e}/4\pi n_0e^2)^{1/2}$, but keeping the original notations, we rewrite the Poisson equation as follows
\begin{eqnarray}
    \partial^2 \varphi/\partial x^2=a \varphi+b \varphi^2+...,
    \label{eq:A3}
\end{eqnarray}
with coefficients $a$ and $b$ dependent on the Mach number and electron-to-ion temperature ratio
\begin{eqnarray}
    a=1-\frac{T_{e}}{2T_{i}}Z'_{r}(M_i),\;\;\;\;b=\frac{1}{2}-\;\left(\frac{T_{e}}{4T_{i}}\right)^2Z^{'''}_{r}(M_i).
    \label{eq:A4}
\end{eqnarray}
Since the observed solitary waves have small amplitudes, $e\varphi/T_{e}\ll 1$, we limit the analysis to the second-order terms on the right-hand side of Eq. (\ref{eq:A3}). For typical electron to ion temperature ratios of $T_{e}/T_{i}\approx 0.1$, we have both $a(M_i)\approx 1$ and $b(M_i)\approx 1/2$ independent of the Mach number (SM), which implies the absence of small-amplitude solitary solutions for Eq. (\ref{eq:A3}). This analysis proves that the observed solitary waves {\it cannot} be described within fluid approximation and, therefore, must be purely kinetic structures.

\section{Width-amplitude relation\label{B}}

In contrast to solitons, whose amplitude and spatial width are strictly related \citep[e.g.,][]{Sagdeev66,Lotko83}, the amplitude and spatial width of electron holes are only related by an inequality \citep[e.g.,][]{Goldman07:prl,Hutchinson17}. The inequality is computed from the Poisson equation written in the form of the integral equation for the distribution function of electrons trapped within an electron hole and results from the positiveness of that distribution. The corresponding inequality obtained by neglecting the ion response reads \citep{Turikov84,Chen04:pre,Chen05:jgr} 
\begin{eqnarray}
    \frac{l^2}{\lambda_{D}^2} \geq \frac{C\;\psi^{1/2}\;{\rm exp}\left(-\psi\right)}{\pi^{1/2} (1-{\rm erf}(\psi^{1/2}))}.
    \label{eq:B0}
\end{eqnarray}
where $\psi=e\varphi_0/T_{e}$ is a normalized amplitude, $l$ is a spatial half-width, and $C$ is a constant weakly dependent on the model used for the electrostatic potential. For example, $C=2(4\;{\rm ln}\;2-1)\approx 3.54$ for $\varphi=\varphi_0\;{\rm exp}(-x^2/2l^2)$ \citep{Chen04:pre,Chen05:jgr}, while for ${\rm sech}^{-2}$ and ${\rm sech}^{-4}$ profiles we have respectively $C\approx 3.46$ and 3.5 \citep{Turikov84}. Note that Eq. (\ref{eq:B0}) was obtained by assuming a Maxwellian distribution of background electrons and also electron hole speed well below electron thermal speed; both assumptions are valid for the electron holes in the Earth's magnetosheath (Figure \ref{Fig:Reduced_VDF}). Eq. (\ref{eq:B0}) was generalized by including the ion response that is applicable only for electron holes, which are at rest in the plasma frame \citep{Chen04:pre,Chen05:jgr}. 

Since the observed electron holes are not at rest in the plasma frame (Figure \ref{Fig:Reduced_VDF}), we need to include a proper ion response. In the small-amplitude limit $e\varphi/T_{e}\ll 1$, the ion response is accurately described by the first two terms on the right-hand side of Eq. (\ref{eq:A2}). The standard computations \citep{Turikov84} result in the following inequality taking into account the proper ion response
\begin{eqnarray}
    \frac{l^2}{\lambda_{D}^2} \geq C\left[\frac{T_e}{T_i}Z^{'}_{r}(M_i)+\frac{\pi^{1/2} (1-{\rm erf}(\psi^{1/2}))}{\psi^{1/2}\;{\rm exp}\left(-\psi\right)}\right]^{-1},
    \label{eq:B1}
\end{eqnarray}
where we will use $C=2(4\;{\rm ln}\;2-1)$. For low Mach numbers, $M_{i}\ll 1$, this inequality reduces to the one obtained by \cite{Chen04:pre,Chen05:jgr}. Note that at sufficiently small amplitudes, the right-hand side of Eq. (\ref{eq:B1}) is dominated by the second term, and the inequality reduces to $l/\lambda_{D}\gtrsim \psi^{1/4}$ that is independent of the ion response, which can be seen in Figure \ref{Fig:histogram}e.

\section*{Data Availability Statement}
\blue{The MMS data used in this paper are publically available at} https://lasp.colorado.edu/mms/sdc/public/. The list of all solitary waves (occurrence times and dates) considered in this paper is available at \cite{Shaikh23:dataset}.

%

\acknowledgments
The work of Z.S. and I.V. was supported by NASA grant No. 80NSSC20K1325 and National Science Foundation grant No. 2026680. The work of I.V. was also supported by NASA grant No. 80NSSC22K1634. J.C.H. was supported by the Los Alamos National Laboratory (LANL) through its Center for Space and Earth Science (CSES). CSES is funded by LANL's Laboratory Directed Research and Development (LDRD) program under project number 20210528CR. I.V. thanks Rachel Wang for valuable comments. We thank the MMS teams for the excellent data. 



%
%
%
%
%
%
%
%
%




\listofchanges

\end{document}